%\mag=\magstephalf
\mag=\magstep1
\pageno=1
\input amstex
%\baselineskip = 0.8 true cm
\documentstyle{amsppt}
\TagsOnRight

\pagewidth{16.5 truecm}
\pageheight{23.5 truecm}
\vcorrection{0.0cm}
\hcorrection{-0.5cm}
%\advance\vsize by -\voffset
%\advance\hsize by -\voffset
\baselineskip = 1.0 true cm
\nologo
\NoRunningHeads
\NoBlackBoxes
\font\twobf=cmbx12

\define \res{\roman {res}}

\define \al{\roman {al}}

\define \Pp{{\roman{P}}}
\define \Qp{{\roman{Q}}}
\define \Py{{P}}
\define \Qy{{Q}}

%\define \tvskip{\vskip 0.5 cm}
\define \tvskip{\vskip 1.0 cm}

\def\fp{\flushpar}

\redefine\qed{\hbox{\vrule height6pt width3pt depth0pt}}

{\centerline{\twobf{On Relations of
Hyperelliptic Weierstrass al Functions}}}

\author
%Tadahiro Ninomiya
Shigeki MATSUTANI
\endauthor
\affil
%8-21-1 Higashi-Linkan Sagamihara 228-0811 Japan
\endaffil
\endtopmatter

%\footnotetext{e-mail:RXB01142\@nifty.ne.jp}

\document
\baselineskip = 0.8 true cm

\centerline{\twobf Abstract }

We study relations of
the Weierstrass's hyperelliptic al-functions
over a non-degenerated hyperelliptic curve $y^2 = f(x)$ of
arbitrary genus $g$ as solutions of sine-Gordon equation
using Weierstrass's local parameters,
which are characterized by two ramified points.
Though the hyperelliptic solutions of the sine-Gordon
equation had already obtained, our derivations of them
is simple; they need only residual computations over
the curve and primitive
matrix computations.

%\subheading{PACS numbers}:
{\centerline{\bf{2000 MSC: 14H05, 14K20, 33E99, 30F30 }}}
{\centerline{\bf{Key Words and Phrases: Hyperelliptic Functions,}}
{\centerline{\bf{sine-Gordon equation, Weierstrass al Functions}}}

%14K20 Analytic theory; abelian integrals and differentials
%33E99 None of the above, but in this section of other special function
%14H05 Algebraic functions; function fields [See also 11R58]
%30F30 Differentials on Riemann surfaces

\baselineskip = 0.8 true cm

%\newpage

\tvskip

{\centerline{\twobf{\S 1. Introduction}}}

The sine-Gordon equation is a famous nonlinear integrable
differential equations.
For a hyperelliptic curve $C_g$ ($y^2=f(x)
 = (x-b_1) \cdots (x-b_{2g+1}) ) $ of genus $g$,
the hyperelliptic solutions of the
sine-Gordon equation were formulated in \cite{Mu 3.241}
 in terms of Riemann theta functions.
In \cite{Mu}, $(U,V,W)$ representation of
symmetric product space of the $g$ curves $Symm^g C_g$
is defined; especially, $U$ is defined by
 $U(z):=(x_1 - z) \cdots (x_g- z)$ a for a point
$((x_1,y_1),\cdots,(x_g,y_g))$ in $Symm^g C_g$.
(In this article, we will denote $U$ by $F(z)$ on later
following the conventions in \cite{Ba1, 2, 3, Ma}.)
Using the relation between $U$
 and the Riemann theta functions
in \cite{Mu 3.113},
the solutions
\cite{Mu 3.241} can be rewritten as,
$$
        \frac{\partial}{\partial t_\Pp}
        \frac{\partial}{\partial t_\Qp}
        \log ( [2\Pp-2\Qp] ) =
        A( [2\Pp-2\Qp ] - [2\Qp-2\Pp ])   , \tag 1-3
$$
where $\Pp$ and $\Qp$ are ramified points of $C_g$,
$A$ is a constant number,
 $[D]$ is a meromorphic function over $\roman{Sym}^g(C_g)$
with a divisor $D$ for each $C_g$
and $t_{\Pp'}$ is a coordinate in the Jacobi variety such that
it is identified with a local parameter at a branch point
$\Pp'$ up to constant. In other words,
for a finite branch point $(b_i,0)$
$U(b_i)$ is identified with $[2(b_i,0)-2\infty]$ up to
constant factor.

In the formulations in \cite{Mu}, local parameters $t_{\Pp'}$
were not concretely treated. In this article,
we will give more explicit
representations of (1-3) using concrete local parameters in
\cite{Ba2, W2, 3} and present simpler derivations of (1-3)
without using any $\theta$-function.
This article is an application of a scheme developed
in \cite{Ma} to the sine-Gordon equation, which is
based upon \cite{Ba3}.

In \cite{W1, W2}, Weierstrass defined $\al$ function  by
$\al_r=\gamma_r\sqrt{U(b_r)}$
using a constant factor $\gamma_r$.
In Theorem 3.1, we will give
$$
\frac{\partial^2}
{\partial u_{1}^{(r)}\partial u_{g}^{(r)}} \log \al_r( u^{(r)})
   = \frac{1}{2}\left( \frac{\al_r^2( u^{(r)})}{\gamma_r^2}
- \frac{f'(b_r)\gamma_r^2}{\al_r^2( u^{(r)})
} \right), \tag 1-4
$$
in terms of a coordinate system $u^{(r)}$'s defined in (2-5).
$\al_r(u)$ has the single order zero at $(b_r,0)$
and a singularity of the single order at $\infty$ as a
function of $x_i \in C_g$.

Further we give another representation in Theorem 4.1
in terms of  $v$'s defined in (2-6) $(a_1:=b_r, a_2:=b_s)$
\cite{W2, 3},
$$
\frac{\partial^2}
{\partial v_1\partial v_2} \log \frac{\al_r( v)}{\al_s(v)}
   = \frac{1}{2(b_r-b_s)}
\left( f'(b_s)\frac{\gamma_s^2\al_r( v)^2}{\gamma_r^2\al_s(v)^2} +
       f'(b_r)\frac{\gamma_r^2\al_s( v)^2}{\gamma_s^2\al_r(v)^2}
          \right). \tag 1-5
$$
The function $\al_s(v)/\al_r(v)$ vanishes with  order one
when $x_i$ is at $(b_s,0)$
 whereas it diverges with  order one if $x_i$ approaches to $(b_r,0)$.
As they were discovered by Weierstrass \cite{W2, 3} and they play
the essential roles in the investigation in \cite{W2, 3} and in \S 4.
Thus we have called them {\it Weierstrass parameters}.

In these proofs, we will use only residual computations
using the data of curve itself without any $\theta$ functions
as the derivation of hypereilliptic solutions of
the modified Korteweg-de Vries equations  in \cite{Ma}.
The curve is sometimes given by an affine equation with special
coefficients. Then it might be important to study the relation between
the properties of line-bundle over the curve and these coefficients.
As (1-4) and (1-5) can be explicitly expressed by data of curve $C_g$,
the author believes that
they have some advantage as relations of special functions.
\tvskip

{\centerline{\twobf{\S 2. Differentials of a Hyperelliptic Curve}}}

In this section, we will give the conventions and notations
 of the hyperelliptic functions in this article.
We denote the set of complex numbers by $\Bbb C$ and
the set of integers by $\Bbb Z$.

\proclaim{\fp 2.1 Hyperelliptic Curve}\rm
We deal with a hyperelliptic curve $C_g$  of genus $g$
$(g>0)$ given by the affine equation,
$$ \split
   y^2 &= f(x) \\  &= \lambda_{2g+1} x^{2g+1} +
\lambda_{2g} x^{2g}+\cdots  +\lambda_2 x^2
+\lambda_1 x+\lambda_0  \\
     &=(x-b_r) h_r(x),\\
\endsplit  \tag 2-1
$$
where $\lambda_{2g+1}\equiv1$ and $\lambda_j$'s are complex
numbers. We use the expressions,
$$
\split
        f(x) &:= (x-b_1)(x-b_2)\cdots(x-b_{2g})(x-b_{2g+1})\\
             &= \Py(x)\Qy(x),\\
        \Py(x) &:= (x-a_1)(x-a_2)\cdots(x-a_{g}),\\
        \Qy(x) &:= (x-c_1)(x-c_2)\cdots(x-c_{g})(x-c),\\
\endsplit \tag 2-2
$$
where $b_j$'s $(b_i=a_i, b_{g+i}=c_i)$ are  complex numbers.
\endproclaim
It is noted that the permutation group acts on these
$\{b_r\}$ and $\{a_r\}$.

\proclaim{\fp 2.2 Definition \cite{Ba1 , Ba2,  W2, 3}}\rm

 \roster
\item
For a point $(x_i, y_i)\in C_g$,
the unnormalized differentials of the first kind are
defined by,
$$   d u^{(r,i)}_1 := \frac{ d x_i}{2y_i}, \quad
      d u^{(r,i)}_2 :=  \frac{(x_i-b_r) d x_i}{2y_i}, \quad \cdots,
     \quad
     d u^{(r,i)}_g :=\frac{(x_i-b_r)^{g-1} d x_i}{2 y_i}.
      \tag 2-3
$$
$$
\split
   d v^{(i)}_1 := \frac{\Py(x_i) d x_i}{2\Py'(a_1)(x_i-a_1)y_i}, \quad
    d v^{(i)}_2 &:= \frac{\Py(x_i) d x_i}{2\Py'(a_2)(x_i-a_2)y_i},
        \quad \cdots,\\ &\quad
      d v^{(i)}_g := \frac{\Py(x_i) d x_i}{2\Py'(a_g)(x_i-a_g)y_i}.
\endsplit
      \tag 2-4
$$

\item
Let us define the Abel maps for $g$-th symmetric product
of the curve $C_g$,
$$
 u^{(r)}:=(u^{(r)}_1,\cdots,u^{(r)}_g)
:\roman{Sym}^g( C_g) \longrightarrow \Bbb C^g,
$$ $$
      \left( u^{(r)}_k((x_1,y_1),\cdots,(x_g,y_g)):= \sum_{i=1}^g
       \int_\infty^{(x_i,y_i)} d u^{(r,i)}_k \right),
      \tag 2-5
$$ $$
v :=(v_1,\cdots,v_g)
:\roman{Sym}^g( C_g) \longrightarrow \Bbb C^g,
$$ $$
      \left( v_k((x_1,y_1),\cdots,(x_g,y_g)):= \sum_{i=1}^g
       \int_\infty^{(x_i,y_i)} d v^{(i)}_k \right).
      \tag 2-6
$$
\endroster
\endproclaim

These coordinates are universal covering of the related Jacobian
$\Cal J$.
The definition (2-6) \cite{Ba2 p.382}
is due to Weierstrass \cite{W2, 3}
and we call (2-6) {\it Weierstrass parameter},
though we choose different constant factor from
the original one  \cite{W2, 3}.
This parameterization is a key of
 the second solutions mentioned in \S 4.

\vskip 0.5 cm
\proclaim {2.3 Definition  }
\rm
\roster
\item Hyperelliptic $al$ function is defined by
\cite{Ba2 p.340, W2, 3},
$$
\roman{al}_r(u) := \gamma_r\sqrt{F(b_r)} , \tag 2-7
$$
where $\gamma_r:=\sqrt{-1/P'(b_r)}$  and
$$
   \split
	F(x)&:= (x-x_1) \cdots (x-x_g)\\
            &=  (x-b_r-x_1+b_r) \cdots (x-b_r-x_g+b_r).
   \endsplit
          \tag 2-8
$$

\endroster

\endproclaim
On the choice of $\gamma_r$,
we will employ the convention of Baker \cite{Ba2} instead of original
one \cite{W2, 3}.
We note that $\al_r$'s have mutually algebraic relations.

For later convenience,
a polynomial associated with $F(x)$ is introduced by
$$
\split
\pi_i^{(r)}(x) &:= \frac{F(x)}{x-x_i}\\
        &=\chi_{i,g-1}^{(r)}(x-b_r)^{g-1}
            +\chi_{i,g-2}^{(r)} (x-b_r)^{g-2}
            +\cdots+\chi_{i,1}^{(r)}(x-b_r)+\chi_{i,0}^{(r)}.\\
\endsplit
$$
Then we have $\chi_{i,g-1}^{(r)}\equiv1$ and
$\chi_{i,0}^{(r)}= F(b_r)/(x_i-b_r)$.
Further we introduce $g\times g$-matrices,
$$
 \Cal W^{(r)} := \pmatrix
     \chi_{1,0}^{(r)} & \chi_{1,1}^{(r)} & \cdots & \chi_{1,g-1}^{(r)}  \\
      \chi_{2,0}^{(r)} & \chi_{2,1}^{(r)} & \cdots & \chi_{2,g-1}^{(r)}  \\
   \vdots & \vdots & \ddots & \vdots  \\
    \chi_{g,0}^{(r)} & \chi_{g,1}^{(r)} & \cdots & \chi_{g,g-1}^{(r)}
     \endpmatrix,\quad
	\Cal Y = \pmatrix
     y_1 & \ & \ & \  \\
      \ & y_2& \ & \   \\
      \ & \ & \ddots   & \   \\
      \ & \ & \ & y_g  \endpmatrix,
$$
$$
\Cal M := \pmatrix
     \dfrac{1}{x_1-a_1} &  \dfrac{1}{x_2-a_1} & \cdots & \dfrac{1}{x_g-a_1} \\
     \dfrac{1}{x_1-a_2} &  \dfrac{1}{x_2-a_2} & \cdots & \dfrac{1}{x_g-a_2} \\
   \vdots & \vdots & \ddots & \vdots  \\
     \dfrac{1}{x_1-a_g} &  \dfrac{1}{x_2-a_g} & \cdots & \dfrac{1}{x_g-a_g}
     \endpmatrix,
$$ $$
	\Cal \Py = \pmatrix
     \sqrt{ \dfrac{\Py(x_1)}{\Qy(x_1)}} & \ & \ & \  \\
      \ &\sqrt{ \dfrac{\Py(x_2)}{\Qy(x_2)}}& \ & \   \\
      \ & \ & \ddots   & \   \\
      \ & \ & \ & \sqrt{ \dfrac{\Py(x_g)}{\Qy(x_g)}}  \endpmatrix,
$$
$$
	\Cal A = \pmatrix
      \Py'(a_1) & \ & \ & \  \\
      \ &  \Py'(a_2)& \ & \   \\
      \ & \ & \ddots   & \   \\
      \ & \ & \ &   \Py'(a_g) \endpmatrix,\quad
	\Cal F^{\prime} = \pmatrix
     F'(x_1) & \ & \ & \  \\
      \ & F'(x_2)& \ & \   \\
      \ & \ & \ddots   & \   \\
      \ & \ & \ & F'(x_{g})  \endpmatrix,\quad
$$
where $F'(x):=d F(x)/d x$.

\proclaim{\fp 2.3 Lemma}\it
For these matrices, following relations hold:
\roster

\item The inverse matrix of $\Cal W^{(r)}$ is given by
$\Cal W^{(r)-1}={{\Cal F}^{(r)\prime}}^{-1}\Cal V^{(r)}$,
where $\Cal V^{(r)}$ is Vandermond matrix,
$$
   \Cal V^{(r)}= \pmatrix 1 & 1 & \cdots & 1 \\
        (x_1-b_r) & (x_2-b_r) & \cdots & (x_g-b_r) \\
       (x_1-b_r)^2 & (x_2-b_r)^2 & \cdots & (x_g-b_r)^2 \\
       \vdots& \vdots &       & \vdots \\
      (x_1-b_r)^{g-1} & (x_2-b_r)^{g-1} & \cdots & (x_g-b_r)^{g-1}
                 \endpmatrix.
$$

\item
$$
\det\Cal M = \frac{(-1)^{g(g-1)/2} \Pp(x_1,\cdots,x_g)\Pp(a_1,\cdots,a_g)}
            {\prod_{k,l}(x_k-a_l)},
$$
where
$$
        \Pp(z_1,\cdots,z_g)  :=\prod_{i<j} (z_i - z_j ).
$$

\item
$$
( \Cal M \Cal \Py)^{-1}\Cal A=
        \left[\left( \frac{2 y_iF(a_j)}{F'(x_i)(a_j-x_i)}
                     \right)_{i,j} \right].\tag 2-9
$$

\endroster
\endproclaim

\demo{Proof}
(1) is obtained by direct computations.
(2) is a well-known result \cite{T}. Since the zero
and singularity in the left hand side give the right hand side as
$$
C \Pp(x_1,\cdots,x_g)\Pp(a_1,\cdots,a_g)/{\prod_{k,l}(x_k-a_l)},
$$
for a certain constant $C$. In order to determine $C$, we multiply
${\prod_{k,l}(x_k-a_l)}$ both sides and let $x_1=a_1$,
$x_2=a_2$, $\cdots$, and $x_g=a_g$. Then $C$ is determined as above.
(3) is obtained by the Laplace formula using the minor determinant
for the inverse matrix.\qed
\enddemo

Then we have following corollary.

\proclaim{\fp 2.5 Corollary}\it
Let $\partial_{u_i}^{(r)}:=\partial/\partial{u_i^{(r)}}$,
$\partial_{v_i}:=\partial/\partial{v_i}$,
and
$\partial_{x_i}:=\partial/\partial{x_i}$.
$$
	\pmatrix \partial_{u_1}^{(r)}\\
                 \partial_{u_2}^{(r)}\\
                 \vdots\\
                 \partial_{u_g}^{(r)}
         \endpmatrix
   =2 \Cal Y \Cal F^{\prime -1}\cdot {}^t\Cal W^{(r)}
	\pmatrix \partial_{x_1}\\
                 \partial_{x_2}\\
                 \vdots\\
                 \partial_{x_g}
         \endpmatrix, \quad
	\pmatrix \partial_{v_1}\\
                 \partial_{v_2}\\
                 \vdots\\
                 \partial_{v_g}
         \endpmatrix
   =2( \Cal M \Cal \Py)^{-1}\Cal A
	\pmatrix \partial_{x_1}\\
                 \partial_{x_2}\\
                 \vdots\\
                 \partial_{x_g}
         \endpmatrix. \tag 2-10
$$
\endproclaim

\tvskip
\centerline{\twobf \S 3. Relations between
Hyperelliptic al Functions $(b_r,\infty)$-type}

In this section, we will give the first  relation of
hyperelliptic al function using the parameters
$u_{1}^{(r)}$ and $u_{g}^{(r)}$ in (2-5).

\proclaim{\fp 3.1 Theorem}\it

$$
\frac{\partial}{\partial u_{1}^{(r)} }
          \frac{\partial}{\partial u_g^{(r)}}\log \al_r =
\frac{1}{2}\left(\frac{\al_r^2}{\gamma_r^2}
- \frac{f'(b_r)\gamma_r^2}{\al_r^2}\right).
      \tag 3-1
$$

\endproclaim

Here we will give a comment on  Theorem 3.1.
Let us fix the parameters $x_2,\cdots,x_g$ and regard
$\al_r$ as a function of a parameter related to $x_1$ over $C_g$.
Then its divisor is $(\al_r) = (b_r,0) - \infty$.
Further by letting $t^2=(x_i-b_r)$ around $(b_r,0)$,
 the definition (2-3) shows,
$$
	d u_{1}^{(r,i)} |_{(b_r,0)} = \frac{2}{\sqrt{f'(b_r)}} d t,
$$
while for $s^2=1/x$ around $\infty$,
$$
	d u_{g}^{(r,i)} |_{(\infty)} = -2 d s.
$$
Hence (3-1) can be regarded as an explicit
 representation of (1-3).

\demo{Proof}
Instead of (3-1), we will prove following
 formula (3-2) in remainder in this section.
$$
\frac{\partial}{\partial u_{1}^{(r)} }
          \frac{\partial}{\partial u_g^{(r)}}\log F(b_r) = F(b_r)
- \frac{f'(b_r)}{F(b_r)}.
      \tag 3-2
$$
The strategy is essentially the same as \cite{Ba3, Ma}.
First we translate the words of the Jacobian into those of
the curves; we rewrite the differentials $u^{(r)}$'s
in terms of the differentials over curves as in (3-3). We count the
residue of an integration and use a combinatorial trick.
Then we will obtain (3-2).

From (2-10), we will express $u^{(r)}$'s by the affine coordinates
$x_i$'s,
$$
	\frac{\partial}{\partial u_g^{(r)} }=
         \sum_{i=1}^g \frac{2y_i}{F'(x_i)} \frac{\partial}{\partial x_i},
$$
$$
	\frac{\partial}{\partial u_{1}^{(r)} }=
        \sum_{i=1}^g \frac{2y_i\chi_{i,0}^{(r)}}{F'(x_i)}
              \frac{\partial}{\partial x_i}
  =F(b_r) \sum_{i=1}^g \frac{2y_i}{(x_i-b_r)F'(x_i)}
              \frac{\partial}{\partial x_i}.
     \tag 3-3
$$
Hence the right hand side of (3-2) becomes
$$
-\frac{\partial^2}{\partial u_1^{(r)}\partial u_g^{(r)} }\log F(b_r)
       =
        F(b_r)
 \sum_{j=1, i=1}^g \frac{2y_j}{( x_i-b_r)^2F'(x_j)}
        \frac{\partial}{\partial x_j}
         \frac{2y_i}{F'(x_i) (x_i-b_r)^2}.
          \tag 3-4
$$

Here we will note the derivative of $F(x)$, which is
shown by direct computations.
$$
	\frac{\partial}{\partial x_k} \left(
   \left[\frac{\partial}{\partial x} F(x)\right]_{x=x_k}\right)
         =\frac{1}{2}
 \left[\frac{\partial^2}{\partial x^2} F(x)\right]_{x=x_k}
         .
$$
Then (3-4) can be written as,
$$
-\frac{\partial}{\partial u_1^{(r)}}
\frac{\partial}{\partial u_g^{(r)} }\log F(b_r)
      =   F(b_r)\sum_{i=1}^g  \frac{1}{F'(x_i)}
             \left[\frac{\partial}{\partial x}\left(
       \frac{f(x)}{(x - b_r) F'(x) }\right) \right]_{x = x_i}
$$
$$
       -  F(b_r)\sum_{k,l, k\neq l}
     \frac{4y_k y_l}{(b_r-x_k)(b_r-x_l) (x_k - x_l)F'(x_k)F'(x_l)}.
$$
The proof of Theorem 3.1 finishes due to the following lemma.
\qed \enddemo

\proclaim{\fp 3.2 Lemma }\it
Following relations hold:
$$
\sum_{k=1}^g \frac{1}{F'(x_k)}
             \left[\frac{\partial}{\partial x}\left(
       \frac{f(x)}{(x - b_r)^2 F'(x)} \right) \right]_{x = x_k}
       = 1 - \frac{f'(b_r)}{F(b_r)^2}
         . \tag 3-5
$$
$$
  \sum_{k,l, k\neq l}
     \frac{2y_k y_l}{(b_r-x_k)(b_r-x_l) (x_k - x_l)F'(x_k)F'(x_l)}
  =0
         . \tag 3-6
$$
\endproclaim

\demo{Proof}:
(3-5) will be proved by the following residual computations:
Let $\partial C_g^o$ be the boundary of a polygon representation
$C_g^o$ of $C_g$,
$$
  \oint_{\partial C_g^o} \frac{f(x)}{(x-b_r)^2F(x)^2} dx =0
 .           \tag 3-7
$$
The divisor of the integrand of (3-7) is given by,
$$
\left(\frac{f(x)}{(x-b_r)^2F(x)^2} dx\right) =
     3\sum_{i=1, b_i \neq b_r}^{2g+1} (b_i,0) -(b_r,0)-
        2\sum_{i=1}^g (x_i,y_i) -2\sum_{i=1}^g (x_i,-y_i) - \infty .
$$
We check these poles:
First we consider the contribution around $\infty$ point.
Noting that the local parameter $t$ at $\infty$ is $x=1/t^2$,
$$
\res_{\infty}\frac{f(x)}{(x-b_r)^2F(x)^2} dx
      = -2.
$$
Since the local parameter $t$ at $(x_k,\pm y_k)$ is
$t=x-x_k$, we have
$$
\res_{(x_k, \pm y_k)}\frac{f(x)}{(x-b_r)^2F(x)^2} dx
      =  \frac{1}{F'(x_k)}
             \left[\frac{\partial}{\partial x}\left(
       \frac{f(x)}{(x - b_r)^2 F'(x) }\right) \right]_{x = x_k}
        .
$$
For each branch point $(b_r,0)$, the local parameter $t$   is
$t^2=x-b_r$ and thus
$$
\res_{(b_r,0)}\frac{f(x)}{(x-b_r)^2F(x)^2} dx
      = 2 \frac{f'(b_r)}{F(b_r)^2}.
$$
By arranging them, we obtain (3-5).

On the other hand, (3-6) can be proved by using a  trick:
 for $i\neq j$,
$$
         \frac{1}{(b_r-x_k)(b_r-x_l) (x_k - x_l)}
  =\frac{1}{ (x_k - x_l)^2}
\left(\frac{1}{(b_r-x_k)}-\frac{1}{(b_r-x_l)}\right).
$$
\qed\enddemo

\tvskip
\centerline{\twobf \S 4. Relations between
Hyperelliptic al Functions: $(a_1,a_2)$-type}

In the previous section, we have a solution with a duality
between a finite ramified point and $\infty$-point.
In this section, we will give a relation between
hyperelliptic $\al$ functions
using the Weierstrass parameter (2-6). The relation
 has a duality between finite ramified points
$(a_r,0)$ and $(a_s,0)$.

\proclaim{\fp 4.1 Theorem }\it
For $r\neq s$, we obtain
$$
\frac{\partial}{\partial v_r }
          \frac{\partial}{\partial v_s}\log \frac{\al_r}{\al_s}
   = \frac{1}{2(a_r-a_s)}
  \left( f'(a_r)\frac{\gamma_r^2\al_s^2}{\gamma_s^2\al_r^2}
  + f'(a_s)\frac{\gamma_s^2\al_r^2}{\gamma_r^2\al_s^2} \right).
      \tag 4-1
$$

\endproclaim

Before we prove it, we will give some comments:
Let us fix the parameter $x_2,\cdots,x_g$ and regard
$\al_r/\al_s(\propto\sqrt{F(a_r)/F(a_s)})$ as a function
of $x_1$ over $C_g$. Then its divisor  is
$(\al_r/\al_s) = (a_r,0) - (a_s,0)$.
By letting $t_r^2=(x_i-a_r)$ around $(a_r,0)$,
infinitesimal value of Weierstrass parameter (2-4) is given,
$$
	d v_{r}^{(i)} |_{(a_r,0)} = \frac{1}{\sqrt{f'(a_r)}} d t_r.
$$
Thus (4-1) is also a concrete expression of (1-3).

\demo{Proof}
Similar to the proof of Theorem 3.1,
let us prove the theorem.
Without loss of generality, we will prove the following
relation instead of (4-1):
$$
\frac{\partial}{\partial v_{1} }
          \frac{\partial}{\partial v_2}\log \frac{F(a_1)}{F(a_2)}
   = \frac{F(a_1)F(a_2)}{(a_1-a_2)}
  \left( \frac{f'(a_1)}{F(a_1)^2}
+  \frac{f'(a_2)}{F(a_2)^2}\right).
      \tag 4-2
$$
From (2-9) and (2-10), the derivative $v$'s are expressed
 by the affine coordinate $x_i$'s,
$$
	\frac{\partial}{\partial v_r}=F(a_r)
         \sum_{j=1}^g \frac{2y_j}{F'(x_j)(x_j-a_r)}
     \frac{\partial}{\partial x_j}.
$$
The right hand side of (4-2) becomes,
$$
\frac{\partial^2}{\partial v_1\partial v_2 }\log \frac{F(a_1)}{F(a_2)}
       =
        F(a_1)
 \sum_{j=1i=1}^g \frac{2y_j}{( x_i-a_1)F'(x_j)} \frac{\partial}{\partial x_j}
         \frac{2y_iF(a_2)}{F'(x_i) (x_i-a_2)}
         \frac{(a_1-a_2)}{(x_i-a_1)(x_i-a_2)}.
          \tag4-3
$$
The right hand side of (4-3) is
$$
         F(a_1)F(a_2)\sum_{i=1}^g  \frac{1}{F'(x_i)}
             \left[\frac{\partial}{\partial x}\left(
       \frac{f(x)(a_2-a_1)}{(x - a_1)^2(x-a_2)^2 F'(x) }\right)
           \right]_{x = x_i}
$$
$$
       -   F(a_1)F(a_2)\sum_{k,l, k\neq l}
     \frac{2y_k y_l(a_2-a_1)}
{F'(x_k)F'(x_l)(x_l-a_1)(x_k-a_2)(x_k-a_1)(x_l-a_2)(x_l-x_k)}.
$$
Then the proof of Theorem 4.1 is completely done due to
the following lemma.
\qed \enddemo

\proclaim{\fp 4.2 Lemma }\it
Following relations hold:
$$
\sum_{i=1}^g  \frac{1}{F'(x_i)}
             \left[\frac{\partial}{\partial x}\left(
       \frac{f(x)}{(x - a_1)^2(x-a_2)^2 F'(x) }\right) \right]_{x = x_i}
       = \frac{1}{(a_1-a_2)^2}
\left(\frac{f'(a_1)}{F(a_1)^2}-\frac{f'(a_1)}{F(a_1)^2}\right)
         . \tag 4-4
$$
$$
  \sum_{k,l, k\neq l}
     \frac{2y_k y_l(a_2-a_1)}
{F'(x_k)F'(x_l)(x_l-a_1)(x_k-a_2)(x_k-a_1)(x_l-a_2)(x_l-x_k)}
  =0
         . \tag 4-5
$$

\endproclaim

\demo{Proof}:
Similar to Lemma 3-2, we consider an integral,
$$
  \oint_{\partial  C_g^o} \frac{f(x)}{(x-a_1)^2(x-a_2)^2F(x)^2} dx =0
 .           \tag 4-6
$$
As the divisor of the integrand of (4-6) is
$$
\split
\left(\frac{f(x)}{(x-a_1)^2(x-a_2)^2F(x)^2} dx\right) &\\
     =3\sum_{i=1, b_i \neq a_1,a_2}^{2g+1} (b_i,0)& -(a_1,0)-(a_2,0)
       - 2\sum_{i=1}^g (x_i,y_i)
       - 2\sum_{i=1}^g (x_i,-y_i) +3 \infty ,\\
\endsplit\tag 4-9
$$
we count residual
contributions from each terms as in the proof of Lemma 3-2
and obtain (4-4). Considering the symmetry,
(4-5) is easily obtained.
\qed\enddemo

\leftline{address:8-21-1 Higashi-Linkan Sagamihara 228-0811 Japan}

\leftline{e-mail:RXB01142\@nifty.ne.jp}
\enddocument